\documentclass{emulateapj}
\usepackage{amsmath}
\usepackage{natbib}
\usepackage{color}
\usepackage{soul}

\newcommand{\rsun}{{R}_{\odot}}
\newcommand{\msun}{{M}_{\odot}}

\newcommand{\Msmin}[1]{M_{#1,\,\mathrm{min}}}

\newcommand{\Jdot}{\dot{J}}
\newcommand{\Mdot}{\dot{M}}
\newcommand{\Porb}{P_{\mathrm{orb}} }
\newcommand{\Psh}{P_{\mathrm{sh}}}
\newcommand{\RL}{R_L}
\newcommand{\MR}{$M$-$R$ }
\newcommand{\tcont}{t_{\mathrm{cont}}}
\newcommand{\Mti}{M_{2,i}}
\newcommand{\Moi}{M_{1,i}}

\newcommand{\Egw}{E_{\mathrm{GW}}}
\newcommand{\Edgw}{\mathcal{E}_{\mathrm{GW}}}
\newcommand{\Lgw}{L_{\mathrm{GW}}}

\newcommand{\ee}[2]{\ensuremath{#1 \times 10^{#2}}}

\newcommand{\pder}[2]{\ensuremath{\frac{\partial \,#1}{\partial#2}}}

\shorttitle{Arbitrarily Degenerate Donors in AM CVn Systems}
\shortauthors{Deloye, Bildsten, \& Nelemans}

\begin{document}

\title{Arbitrarily Degenerate Helium White Dwarfs as Donors in AM CVn Binaries} 
\author{Christopher J. Deloye}
\affil{Department of Physics, Broida Hall, University of California, Santa Barbara, CA 93106 \altaffilmark{1}}
\email{cjdeloye@northwestern.edu}
\author{Lars Bildsten}
\affil{Kavli Institute for Theoretical Physics and Department of Physics, Kohn Hall,University of California, Santa Barbara, CA 93106}
\email{bildsten@kitp.ucsb.edu}
\and
\author{Gijs Nelemans}
\affil{Department of Astrophysics, Radboud University Nijmegen, P. O. Box 1910, NL-6500 GL Nijmegen, The Netherlands}
\email{nelemans@astro.ru.nl}
\altaffiltext{1}{currently at Department of Physics \& Astronomy, Northwestern University, 2131 Tech Drive, Evanston, IL 60208}

\begin{abstract}
We apply the \citet{delo03} isentropic models for donors in ultracompact low-mass X-ray binaries  to the AM CVn population of ultracompact, interacting binaries.  The mass-radius relations of these systems' donors in the mass range of interest ($M_2<0.1 \msun$) are not single-valued, but parameterized by the donor's specific entropy.  This produces a range in the relationships between system observables, such as orbital period, $\Porb$, and mass transfer rate, $\Mdot$. For a reasonable range in donor specific entropy,  $\Mdot$ can range over several orders of magnitude at fixed $\Porb$.  We determine the unique relation between $\Mdot$ and $M_2$ in the AM CVn systems with known donor to accretor mass ratios, $q=M_2/M_1$. We use structural arguments, as well as each system's photometric behavior, to place limits on $\Mdot$ and $M_2$ in each.  Most systems allow a factor of about 3 variation in $\Mdot$, although V803 Cen, if the current estimates of its $q$ are accurate, is an exception and must have  $M_2 \approx 0.02 \msun$ and $\Mdot \approx 10^{-10} \msun$ yr$^{-1}$.  Our donor models also constrain each donor's core temperature, $T_c$, range and correlate $T_c$ with $M_2$. We examine how variations in donor specific entropy across the  white dwarf family \citep{nele01a} of AM CVn systems affects this population's current galactic distribution.  Allowing for donors that are not fully degenerate produces a shift in systems towards longer $\Porb$ and higher $\Mdot$ increasing the parameter space in which these systems can be found.  This shift increases the fraction of systems whose $\Porb$ is long enough that their gravity wave (GW) signal is obscured by the background of detached double white dwarf binaries that dominate the GW spectrum below a frequency $\approx 2$ mHz.   
\end{abstract}


\section{Introduction \label{sec:intro}}
The AM CVn class of variable stars are He-rich objects which show a striking absence of H and photometric and spectroscopic variations with periods of $\approx$ 300-4000 s \cite[see][for an overview]{warn95}. These periods appear to be orbital and the commonly held model for these systems is that of a mass-transferring binary with a low-mass He white dwarf (WD) donor and a C/O or He WD accretor \citep{pacz67, faul72}.  This model is supported by several lines of evidence: the spectra are dominated by double peaked lines \citep[e.g.,][]{marsh95, groot01}, implying the presence of an accretion disk \citep{nath81,nass01}, rapid photometric flickering is observed, suggesting ongoing mass transfer \citep[e.g.][]{warn72a, warn72b}, and, in several systems, the spectral line profiles vary with a stable, regular period coincident with one of the primary photometric periods, indicating this is the system's $\Porb$ \citep{nath81, nele01b}.  These systems' large observed spectral line broadening requires a compact accretor \citep{nath81,patt93, ruiz01, groot01} while their weak X-ray emission indicates the accretor is a WD \citep{nath81,ruiz01,ulla95}.    

Because AM CVn systems are undergoing mass transfer, we can constrain the donor's structure from the orbital period, $\Porb$. For stable mass transfer, the donor's radius, $R_2$, must equal its Roche lobe's radius, $\RL$, which can be approximated as
\begin{equation}
\RL \approx 0.46 a \left(\frac{M_2}{M_1+M_2}\right)^{1/3}\,,
\label{eq:RL}
\end{equation} 
when $q \equiv M_2/M_1 < 0.8$  \citep{pacz67}.  Here $M_1$ and $M_2$ are the masses of the accretor and donor respectively and $a$ is the orbital separation (throughout, we assume that the orbit is circular).  Combining $\RL = R_2$ with Kepler's third law leads to the  period-mean density relation,  
\begin{equation}
\Porb \simeq 101\,\mathrm{s} \left(\frac{R_2}{0.01 \rsun}\right)^{3/2} \left(\frac{0.1 \msun}{M_2}\right)^{1/2}\,.
\label{eq:pdrel}
\end{equation}
In analyzing and modeling AM CVn systems, it is customary to specify a mass-radius, $M$-$R$, relation for the donor producing a one-to-one relation between $M_2$ and $\Porb$.

Two such \MR relations have been utilized extensively, each motivated by a different AM CVn formation channel. In both of these channels, several common envelope (CE) episodes \citep[see, e.g.,][]{pacz76, yung93} are involved in producing a close binary pair consisting of a degenerate WD accretor and a low-mass He core remnant. The latter may eventually becomes the donor when and if gravity wave angular momentum losses bring the pair into contact.  In the first channel, which we will refer to as the WD-channel, this He object  never ignited after the end of the CE phase, and the donor makes contact as a (possibly semi-) degenerate He WD. In this case, the donor is typically modeled using a $T=0$ \MR relation such as that of \citet{zapo69} \citep[e.g.,][]{nele01a}. In the other channel, which we refer to as the He-star channel, the He object is able to ignite He and makes contact as a He-burning star \citep{savo86, tutu89,ergma90,nele01a}. For this channel, \citet{nele01a} used a fit to the evolution track of model 1.1 from \citet{tutu89} to obtain a donor \MR relation, while \citet{warn95} has also quoted a \MR fit to the \citet{savo86} evolutionary calculation. In addition to these two channels, \cite{pods02,pods03} have recently proposed a scenario involving stable mass transfer from a H-depleted main sequence star that could also produce AM CVn systems.  The potential contribution to the AM CVn population from this channel has not be systematically explored nor have \MR relations for the donors produced in this channel been reported.

These sets of \MR relations have been used extensively to approximate the properties of known systems \citep[e.g., ][]{warn95,prov97,marsh99,nass01,elkhoury00,nele01b,wood02} and in AM CVn population synthesis studies \citep{nele01a,farm03}. However, the use of discrete, single-valued \MR relations limits what we can learn about these systems.  For example, they do not provide a means of inferring internal donor properties  from observations.  Nor do they allow us to model the effects of the expected differences between systems formed in a given channel.  For example, in the WD-channel {there should be variations in the donor's entropy at initial contact} \citep{tutukov96,bild02}. In the He-star channel, {the amount of He burned to C} impacts the system's minimum $\Porb$ and the donor's subsequent evolution \citep{ergma90}.  A discrete set of \MR relations does not allow us to model the impact of these differences on the AM CVn population nor to interpret potential observational signs of such variations.

Progress on these fronts has been hampered by a lack of models in the mass range $(M_2<0.1 \msun$) applicable to the AM CVn donors.  Our recent calculation of models for low-mass donors in ultracompact low-mass X-ray binaries \citep[][hereafter DB03]{delo03} now provides {a set of models} continuously parameterized {by the donor's specific entropy} in this regime. {In contrast to the above \MR relations, our model set's \MR relation is not single-valued, therefore allowing us to make connections between the macroscopic and internal properties of the donor.} In this paper we apply the DB03 models to the AM CVn binaries.  

In \S 2, we summarize the details of our models and compare their continuous \MR relations to the discrete set of \MR relations {above}.  We highlight how a range in the donor's physical parameters translates into a range in observables, focusing on the relation between the secular mass transfer rate, $\Mdot$, and $\Porb$.  We  apply our models to the known AM CVn binaries in \S 3. For systems where $q=M_2/M_1$ is known or inferred, there is a single-valued relation between $\Mdot$ and $M_2$. Making reasonable assumptions about the nature of the donor and accretor, we place limits on $\Mdot$ and $M_2$. Our models provide a connection between these constraints and limits on the donor's specific entropy, \emph{the first such constraints on the AM CVn donors}.  With the current observations, this does not yet provide a strong constraint on donor properties (except possibly for the case of V803 Cen which must have $T_c<10^6$ K given its assumed $q$). With stronger observational limits,  {we may be able to} constrain the donor's prior evolution and  {possibly} provide a means of determining  {individual system's} formation channel. In \S 4, we use our models to explore how a range in donor specific entropy at contact in WD-channel AM CVn systems  alters the $\Mdot,\,\Porb$ distribution from that calculated assuming a $T=0$ \MR relation.  A sizable fraction of systems in this population make contact within several 100 Myr or so of the end of the last CE phase \citep{tutukov96}, not allowing the donors to cool much before mass transfer begins.  These high specific entropy donors follow tracks that differ from the $T=0$ evolution track. The consequence is a population that occupies a larger region of $\Mdot$-$\Porb$ parameter space. {In general, higher entropy donors follow tracks shifted towards longer $\Porb$ relative to fully degenerate donors.}  We also discuss how this calculation impacts the gravity wave, GW, signal produced by this population. In particular, by the shifting of systems with hotter donors to longer $\Porb$, the number of AM CVn systems, relative to a population with only fully degenerate donors, that will be undetectable behind the detached WD-WD binary GW background increases.  Finally, in \S 5 we summarize our results and discuss future directions for study.

\section{Hot Low-Mass Helium Donors: Implications for Interacting Binaries \label{sec:genprops}}
Details concerning the calculation of our low-mass He models can be found in \citet{delo03}.  Briefly, we integrate hydrostatic balance using a modern equation of state (EOS) \citep{chab98, pote00} that includes realistic treatment of Coulomb contributions, which are important at the low densities of these low-mass objects. Knowing the internal thermal profile of an object is impossible without following a given evolutionary scenario, so we assume the limiting case of an adiabatic internal profile; see DB03 for further justification.  We calculate models for $2 \leq \log(T_c/\mathrm{K}) \leq 7.9$ and $2.0-2.3 \leq \log(\rho_c/\mathrm{g\, cm^{-3}}) \leq 6.6$ (the lower limit depending on $T_c$) to produce a set of models parameterized by specific entropy.

\begin{figure}
\plotone{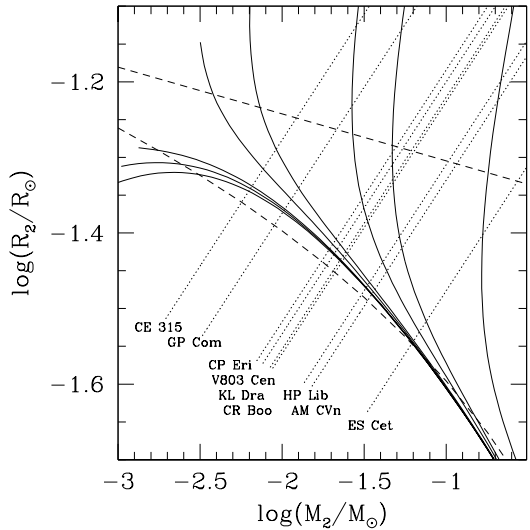}
\caption{A comparison between the He WD models of DB03 and the \MR relations currently used in the AM CVn literature.  The solid lines show a set of isotherms for the DB03 He WD models with $\log(T_c/K)$ between 4.0 and 7.5 at intervals of $\Delta \log(T_c/K)$ =0.5. The lower dashed line shows the $T=0$ \citet{zapo69} He WD relation.  The upper dashed line shows the \MR relations for the He-star channel used by \citet{nele01a}.  The dotted lines show the $\RL$ filling solutions at the $\Porb$ of the indicated systems (see Table \ref{tab:systems} for these system's parameter). \label{fig:hemr}}
\end{figure}

We show in Figure \ref{fig:hemr} the \MR relations for our He models by the solid lines.  Each line shows a sequence with the same $T_c$ and terminates with models that fill their $\RL$ at $\Porb \approx 170$ min. The coldest isotherm ($T_c=10^4$ K) is equivalent to the $T=0$ \MR relation.  The difference in this curve and the \citet{zapo69} relation as fit by \citet{nele01a} (the lower dashed line) is due to the more simplified treatment of Coulomb physics in the \citet{zapo69} EOS as compared to that of \citet{chab98} and \citet{pote00}. The turn-over in the \MR relations seen at low $T_c$ is due to Coulomb interactions becoming the dominant source of binding in the WD.  As $T_c$ increases, thermal contributions to the EOS at low density eventually dominate Coulomb contributions and the \MR relation no longer turns over (for He objects, this occurs for $T_c \approx 10^5$ K, DB03).  For higher $T_c$, there is a minimum mass below which equilibrium models do not exist \citep{cox68, cox64, hans71, rapp84} that is related to the transition between thermal pressure and degeneracy pressure dominating the object's structure \citep{delo03}. Correspondingly, for each isotherm, there are 2 branches to the \MR relation: an upper branch where the object is thermally supported and a lower branch where degenerate electrons provide the dominant pressure support.  For a fixed  $M_2$, on the lower branch, $R_2$ increases with $T_c$, while on the upper branch, $R_2$ decreases with increasing $T_c$ (a fact related to the negative heat capacity of thermally supported stars).  For comparison, the upper dashed line shows the \citet{nele01a} fit to the He-star \MR track of \citet{tutu89} used in the \citet{nele01a} population synthesis calculation and elsewhere.

The dotted lines in Figure \ref{fig:hemr} show \MR relations required for the donor to fill it Roche lobe (equation \ref{eq:pdrel}) at the $\Porb$ for the indicated AM CVn systems. At fixed $\Porb$,  the minimum $M_2$ a donor can have and satisfy this constraint is set by the intersection of each dotted line with the $T=0$ \MR relation. For $M_2$ greater than this minimum, there is a continuous set of $\RL$ filling solutions parameterized by the donor's entropy.  \emph{{Thus}, at fixed $\Porb$, the donor's specific entropy determines $M_2$}. On the lower branch of the \MR relations, hotter donors must be more massive as seen along the dotted lines in Figure \ref{fig:hemr}.

\begin{figure}
\plotone{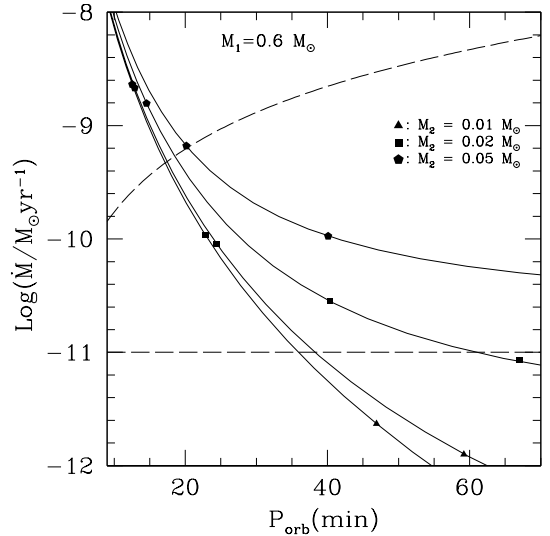}
\caption{The relation between $\Mdot$ and $\Porb$ for fixed $M_1 = 0.6 \msun$ as produced by our He WD models.  The solid lines show this relation at constant $T_c = 10^4,\, 10^6,\,\ee{5}{6},$ and $10^7$  K (from bottom to top).  The symbols indicate where along each isotherm $M_2 =$ $0.01 \msun$ (triangles), $0.02 \msun$ (squares), and $0.05 \msun$ (pentagons). Along several hotter isotherms, the same mass is indicated at two points, a situation resulting from the existence of both upper and lower branch models at these $T_c$. In these cases, for the shorter $\Porb$ point, the donor is on the lower \MR branch, while for the longer $\Porb$ point, the donor is on the upper \MR branch. The \citet{tsug97} He disk stability criteria (with $q=0.05$) are shown by the dashed lines. At longer $\Porb$, allowing for hot donors produces a several orders of magnitude range in $\Mdot$. \label{fig:mdotp_isoT}}
\end{figure}

\begin{figure}
\plotone{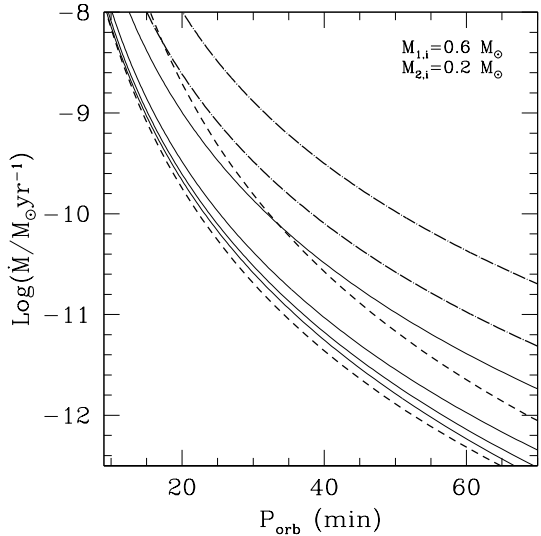}
\caption{The relation between $\Mdot$ and $\Porb$ along \emph{evolutionary} tracks.  Each line shows evolution from an initial configuration of $M_{2,\mathrm{i}}=0.2 \msun$ and $M_{1,\mathrm{i}}=0.6 \msun$. The solid lines show the adiabatic evolution of our He WD models initially with $T_c=10^4,\,10^7,\,\ee{3}{7}$, and $\ee{6}{7}$ K (left to right) on the \emph{lower} branch of the \MR relations.  The dot-dashed lines show the same for initial $T_c=\ee{6}{7}$ and $\ee{5}{7}$ K (again, left to right) on the \emph{upper} branch.  The dashed lines show the evolution using the \citet{nele01a} fits for the $T=0$ (lower) and He-star (upper) \MR relations. %
\label{fig:mdotp_adiab}}
\end{figure}

The fact that $M_2$ and $R_2$ can take on a range of values  produces a range in binary parameters that is related to the donor's specific entropy.  To illustrate this, consider $\Mdot$, which, assuming conservative mass transfer, is given by
\begin{equation}
\frac{\Mdot}{M_2} =2 \Big\vert \frac{\Jdot}{J}\Big\vert \frac{1}{n+5/3 - 2 q}\,.
\label{eq:mdot}
\end{equation}
Here $J$ is the orbital angular momentum, $\Jdot<0$ the angular momentum loss rate from GW emission \citep{land62} and
\begin{equation}
n \equiv \pder{\ln R_2}{\ln M_2}\,.
\label{eq:n}
\end{equation}
Throughout this paper we assume that the donor responds adiabatically so that $n$ is calculated along a well-defined adiabat.  This is equivalent to assuming that no specific entropy is generated or lost in the donor and that $\Mdot$ is high enough that the orbital evolution timescale is shorter than the donor's cooling time.  This latter assumption seems well justified for ultracompact binaries \citep{bild02}, {but see \S 4 for a further discussion of the validity of this assumption}. With equations (\ref{eq:pdrel}) and (\ref{eq:mdot}), we relate $\Mdot$ to $\Porb$ based on $M_2$, $R_2$, $n${, and assuming some value for $M_1$}.

First consider this relation assuming there is no restriction on $T_c$ (other than it is low enough so that He burning cannot occur) at any $\Porb$.  In this case, $\Mdot$ can vary by as much a factor of 100 or more at fixed $\Porb$.  This is illustrated in Figure \ref{fig:mdotp_isoT} by the solid lines, which show the $\Mdot$-$\Porb$ relation along lines of constant donor $T_c$ assuming a fixed $M_1=0.6 \msun$ ($\Mdot$ varies a small amount for $M_1= 0.4-1.4 \msun$).  It should be noted that these lines do not represent an evolutionary sequence, since $n$ along an adiabat is used to calculate $\Mdot$.  Instead, they show the parameters needed for a system to have some $\Mdot$ at a given $\Porb$.   The dashed lines show the upper and lower thermal stability criteria for a He disk as determined by \citet{tsug97}. Above the upper line, the disk is in a high-luminosity stable state; below the lower line, the disk is in a low-luminosity stable state.  For systems located between these two lines, the disk is expected to exhibit periodic outbursts (see the discussion in \S 3).  A population of AM CVn binaries with variable evolution histories should be expected to have a range in $\Mdot$ at fixed $\Porb$, with $\Mdot$ providing a diagnostic of $M_2$ and $T_c$ in individual systems.  Additionally, the rate at which a given AM CVn systems evolves in $M_2$ and in $\Porb$ also varies with the donor's specific entropy (DB03).  

The plausible range in $T_c$ for a given $\Porb$ depends on the formation channel and past evolution so that the range in $\Mdot$ at longer $\Porb$ will depend on the properties of the progenitor population. We show this by considering adiabatic evolution from the minimum $\Porb$.  The donor in this case follows an adiabat set by its specific entropy at contact. The core temperature decreases with $M_2$ as the donor expands along the adiabat. We show a few of the resulting $\Mdot$-$\Porb$ evolution tracks in Figure \ref{fig:mdotp_adiab} calculated from the initial conditions $M_{1,\mathrm{i}}=0.6 \msun$ $M_{2,\mathrm{i}}=0.2 \msun$.  The solid lines show tracks for our He models starting on the lower, degenerate, branch of the \MR relations, each line corresponding to a different initial $T_c$ (between $10^4$ and $\ee{6}{7}$ K).  The dot-dashed show hot models initially on the upper branch.  The dashed lines show the evolution along the He-star (upper dashed line) and $T=0$ (lower dashed line) tracks of \citet{nele01a}. Donors from the WD-channel are much less likely to make contact while on the upper branch (due to their rapid cooling while hot).  Objects that do make contact while on the upper branch will follow tracks similar to the dot-dashed lines at much higher $\Mdot$ than the those with $T=0$ donors (shown by the lowest solid line Figure \ref{fig:mdotp_adiab}). The distribution of $\Mdot$ at a specific $\Porb$ depends on the distribution of the donors' specific entropy at contact (assuming donors evolve adiabatically). We discuss in \S 4 the properties of this distribution in the context of the WD-channel population synthesis model of \citet{nele01a}.

The AM CVns produced through the He-star channel will have, on average, donors with higher specific entropy than those from the WD-channel, thus populating a region in Figure \ref{fig:mdotp_adiab} above the $T=0$ track. The track produced by the \citet{nele01a} fit to a He-star evolution (the upper dashed line in Figure \ref{fig:mdotp_adiab}) provides one such example. The difference in the slope of this track and the DB03 adiabats may be due to several causes---the donor cooling over the course of the \citet{tutu89} calculation is one possibility, differences in the internal structure of the models another---and indicates that the approximations inherent in our models and in calculating their evolution are only a first step towards a more realistic understanding of compact binary evolution.  We will discuss some future directions for improvements to our calculations below.

\section{Application to the Known AM CVn Systems \label{sec:specprops}}
\begin{figure}
\plotone{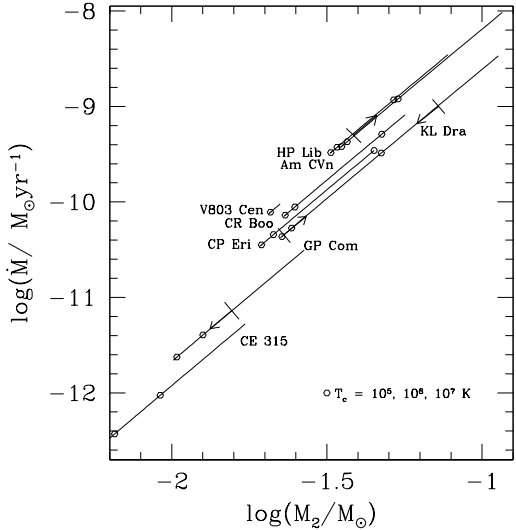}
\caption{For the systems listed in Table \ref{tab:systems}, the solid lines show the relation between $M_2$ and $\Mdot$ based on $\Porb$ and $q=M_2/M_1$.  The lower limit on each curve is set by  our $T=0$ \MR relations, the upper limit by the requirement that $M_1 \leq 1.4 \msun$.  For some systems, the accretion disk's thermal stability, inferred from the system's photometric behavior, further constrains $\Mdot$. The bold lines with arrows indicate such upper and lower limits where applicable (for clarity we show only one arrow for AM CVn and HP Lib which share a lower disk-set $\Mdot$ limit). The circles along each solid line indicate where our donor models have $T_c =10^5,\,10^6,\,$ and $10^7$ K (from left to right). \label{fig:mdot_m2}}
\end{figure}

We now apply our models to the AM CVn systems and determine the constraints that observations place on the binary and the donor. Each binary has a range in $M_2$ over which the donor can fill its Roche Lobe, leading to a range of possible binary parameter. For most AM CVn systems, $q=M_2/M_1$ has either been spectroscopically measured \citep{nath81, marsh99} or can be inferred from $\Porb$ and the so-called superhump period observed in many of the AM CVn systems \citep{white88, hiro90,hiro93, ichi93, solh98, skil99, prov97, patt97, patt02}. In all cases $q<0.1$ \citep[see, e.g.,][as well as Table \ref{tab:systems}]{woudt03} .  We use these $q$ to obtain a unique relation between $\Mdot$ and $M_2$ for each AM CVn system. These relations, calculated using equation (\ref{eq:mdot}), are shown in Figure \ref{fig:mdot_m2} by the solid lines.

The minimum value of $M_2$ for each system is set by the intersection of the DB03 $T=0$ \MR relation with the Roche-lobe filling solution to equation (\ref{eq:pdrel}) at the appropriate $\Porb$ {(see Figure \ref{fig:hemr})}.  The upper limit on $M_2$ is set by $q$ and the requirement that $M_1<1.4 \msun$, the Chandrasekhar mass. The relations described by the solid lines only depend on the donor model through $n$ in equation (\ref{eq:mdot}).  For $M_2<0.1 \msun$, DB03 show that $n$  varies by $\approx 5\%$ between He models with $T_c=0$ and $T_c=10^7$ K.  Hence, these $\Mdot$-$M_2$ relations are only weakly dependent on the donor model.  Between these limits,  each system has a potential spread in $M_2$ of a factor of about 3. The exception is V803 Cen, whose very small $q$ sets a minimum whose accretor mass of $\Msmin{1} \approx 1.3 \msun$, so that $M_2$ has an extremely narrow allowed range: $0.021 \msun \leq M_2 \leq 0.022 \msun$.  {However, since there is some uncertainty in the identification of $\Porb$ and the superhump period in V803 Cen \citep{kato04}, and therefore in the derived $q$,  for this system, these constraints should be considered tentative.} For the other systems, roughly half (HP Lib, CR Boo, CP Eri, and GP Com) have $\Msmin{1} > 0.5 \msun$, indicating that the accretor could be a C/O or Hybrid \citep{iben85} WD. However, since $M_1$ may have grown as much as $\approx 0.1-0.2 \msun$ during the AM CVn phase, the possibility the accretor in these systems is a He WD is not ruled out.

\begin{deluxetable*}{l c c c c c c c l}
\tablecaption{AM CVn System Parameters with Limits Set by Donor and Accretor Stellar Structure \label{tab:systems}}
\tablehead{\colhead{System}&\colhead{$\Porb$ (s)}&\colhead{$\Psh$ (s)}&\colhead{$q$}&\colhead{$\left(\frac{M_{2, \mathrm{min}}}{\msun}\right)$\tablenotemark{a}}&\colhead{$\left(\frac{M_{1, \mathrm{min}}}{\msun}\right)$}&\colhead{$\log\left(\frac{\Mdot}{\msun\,\mathrm{yr^{-1}}}\right)$}&\colhead{$\log\left(\frac{T_{c,\mathrm{max}}}{\mathrm{K}}\right)$}&\colhead{References}}
\startdata
AM CVn&1029&1051&0.087&0.035& 0.40& -9.43 -- -8.01 & 7.48&\citet{nele01b}\\*&&&&&&&&\citet{skil99}\\*&&&&&&&&\citet{solh98}\\
HP Lib&1103&1119&0.056&0.032&0.625&-9.49 -- -8.46 & 7.28&\citet{patt02}\\
CR Boo&1471&1487&0.042&0.023&0.547&-10.12 -- -9.09& 7.10&\citet{prov97}\\*&&&&&&&&\citet{wood02}\\*&&&&&&&&\citet{patt97}\\
KL Dra&1502&1533&0.081&0.022 &0.277&-10.37 -- -8.47& 7.40 &\citet{wood02}\\
V803 Cen&1612&1619&0.016&0.021&1.29&-10.12 -- -10.02 & 5.95 &\citet{patt00}\\
CP Eri&1701&1716&0.03&0.019&0.641&-10.46 -- -9.46& 6.98&\citet{patt01}\\
GP Com&2794&\nodata&0.022\tablenotemark{b}& 0.010&0.46&-11.66 -- -10.50& 6.70&\citet{marsh99}\\
CE-315&$3906\pm42$&\nodata&0.0125\tablenotemark{b}&0.006&0.48&-12.47 -- -11.28 & 6.48&\citet{ruiz01}\\*&&&&&&&& D. Steeghs, priv. comm.\\
\enddata
\tablenotetext{a}{\phd Lower limits on $M_2$, $M_1$, $\Mdot$, and $T_c$ are set by DB03 $T=0$ He \MR relations, upper limits by $M_1<1.4 \msun$.} 
\tablenotetext{b}{\phd Determined from spectroscopy.}
\end{deluxetable*}

\begin{deluxetable*}{l c c c c}
\tablecaption{Limits on AM CVn System Parameters set by He Disk Stability\tablenotemark{a} \label{tab:systems_disk}}
\tablehead{\colhead{System}&\colhead{$\left(\frac{M_{2, \mathrm{min}}}{\msun}\right)$\tablenotemark{a}}&\colhead{$\left(\frac{M_{1, \mathrm{min}}}{\msun}\right)$}&\colhead{$\log\left(\frac{\Mdot}{\msun\,\mathrm{yr^{-1}}}\right)$}&\colhead{$\log\left(\frac{T_c}{\mathrm{K}}\right)$}}
\startdata
AM CVn &0.039&0.45&-9.30 -- -8.01\tablenotemark{b}&6.38 -- 7.48\tablenotemark{b}\\
HP Lib&0.038&0.677&-9.30-- -8.46\tablenotemark{b}&6.52 -- 7.28\tablenotemark{b}\\
KL Dra&0.023&0.282&-10.34 -- -9.00& 5.33 --7.22\\
GP Com&\nodata&\nodata&-11.66\tablenotemark{b} -- -11.14 & $<$ 6.32\\
\enddata
\tablenotetext{a}{\phd Calculated using \citet{tsug97} He disk thermal stability criteria.}
\tablenotetext{b}{\phd Limit set by donor or accretor structural limits; see Table \ref{tab:systems}.}
\end{deluxetable*}

Observations place further constraints on $M_2$ in some of these systems. There is an observed correlation between the photometric behavior of AM CVn systems and their $\Porb$.  The short period systems, AM CVn and HP Lib, are bright and do not exhibit large amplitude variability \citep{warn95,solh98, skil99}.  The intermediate period objects---CR Boo, KL Dra, V803 Cen, and CP Eri---exhibit large amplitude variability \citep[$\approx$ 3-4 magnitude;][]{wood87, schw98, wood02, odon89,patt00,abbott92},  while the long period systems, GP Com and CE 315, are consistently faint and again do not exhibit large amplitude brightening \citep{rich73, ruiz01, woudt01}.  This behavior is believed to be explained with a thermal-tidal He accretion disk instability model \citep{smak83, tsug97} that relates $\Mdot$ to these different observed photometric regimes.  Taking the \citet{tsug97} results, we use their stability criteria to place limits on the allowed $\Mdot$ in the AM CVn systems based on the photometric behavior of each. In some cases, disk stability gives stronger constraints than those above and we show these additional limits in Figure \ref{fig:mdot_m2} as upper and lower limits on $\Mdot$.  

From our He WD models, we can also determine $T_c$ as a function of $M_2$ in each system.  We indicate the models with $T_c=10^5$, $10^6$, and $10^7$ K by open circles along each system's $\Mdot$-$M_2$ relation in Figure \ref{fig:mdot_m2}.  We summarize the limits placed on these systems from donor and accretor structural constraints, along with other system properties, in Table \ref{tab:systems}. For those systems in which He disk stability arguments provide stronger limits, we provide these limits in Table \ref{tab:systems_disk}.  \emph{These represent the first limits placed on the donor's specific entropy in AM CVn systems}.  Again, given current observations, most of these limits are not strong.  However, the lower limits on AM CVn and HP Lib set by disk stability arguments indicate their donors are far from being $T=0$ objects.  In V803 Cen, the donor has a maximum $T_c\approx 10^6$ K, placing it on the degenerate branch of the \MR relations. Perhaps this provides evidence that V803 Cen formed through the WD-channel; however, a better understanding of the range in the donor's initial specific entropy in each formation channel is needed to support such a claim. Combined with stronger observational limits on other AM CVn systems, connections between the formation channel and the donor's specific entropy could provide one tool for determining a system's prior evolution.     

\section{AM CVn Systems formed through the WD Channel \label{sec:evolimpact}}
We have shown that the AM CVn binaries can occupy a larger region of $\Mdot$-$\Porb$ parameter space than previously thought.  There remains the question, however, of how much of this parameter space they actually occupy given reasonable evolution scenarios. In the WD channel, variations in $M_2$ and in the time the donor has to cool before it makes contact, $\tcont$,  produces donors with different initial specific entropies. Figure \ref{fig:mdotp_adiab} showed how this can lead to a wide range of evolutionary tracks. In the He-star channel, the donor initiates mass transfer as a He-burning star and its structure depends on $M_2$ and the amount of He burned to C/O.  If the composition was fixed, then there would be a single \MR relation (the He-burning main-sequence) and all systems would have the same minimum $\Porb$ and the same subsequent evolution, modulo the  variations due to a range in $M_1$ {(up to a factor of 5)}.  However, variations in $\tcont$ impact how much He is processed before nuclear reactions are quenched by mass loss,  affecting the donor's \MR relation, the binary's minimum $\Porb$, and producing variation in the subsequent evolution during the AM CVn phase. Our current models cannot accommodate varying internal composition and so we will only address variability within the WD-channel.   We refer the reader to \citet{pods02, pods03} and \citet{nels03} for discussions of the evolutionary variations in AM CVns and other ultracompact binaries formed through the H-depleted main sequence star channel. 

We now show how a range in $\tcont$ affects the $\Porb$-$\Mdot$ distribution of WD-channel AM CVn systems. We begin with the population synthesis data from \citet{nele01a}.  We refer the reader to \citet{nele01a} and references therein for the details of their model.  The information needed for our calculation are the initial accretor and donor masses, $\Moi$ and $\Mti$, at the beginning of the AM CVn stage,  $\tcont$,  and the galaxy's age when the system begins the AM CVn stage.  We use $\tcont$ to determine the donor's specific entropy at contact using existing He WD evolution calculations.  We evolve each system adiabatically from contact assuming conservative mass transfer driven by GW emission up to the present age of the galaxy (taken to be 10 Gyrs) in agreement with \citet{nele01a}.  We removed systems where $M_1$ grew larger than 1.4 $\msun$ during this evolution.

\subsection{Determining the Initial Donor Model \label{sec:initial_donor}}
The donor starts out as the core of a red giant branch star which initiates unstable mass transfer and a CE phase.  The state of the outer regions of the donor after the CE phase ends (e.g., how much of a H envelope is left) will depend on the uncertain details of CE evolution, but the central conditions of the donor will be largely unchanged from their state at the start of the CE phase. Afterwards, the donor will settle towards a cooling He WD configuration on its thermal time ($\approx 10^6$-$10^7$ yr).  In the meantime, it cools and contracts until it makes contact at a time $\tcont$ later.  The system's evolution through these  phases determines the specific entropy of the donor at contact, and the distribution of the donor's specific entropy at contact determines the number of systems evolving along a specified range of adiabatic tracks (see Figure \ref{fig:mdotp_adiab}) during the subsequent AM CVn phase and, hence, the resulting population distribution. {For donors with $M_2 \approx 0.01-0.03 \msun$, as is appropriate for the known AM CVns, it is the specific entropy of the inner $\approx$ 10\% of the initial donor mass that matters for current the evolution.}  We thus emphasize a proper accounting of the donor's central conditions throughout the pre-contact evolution to ensure we calculate a realistic population distribution for $\Porb \gtrsim 20$ min.   

To determine the initial donor model, we begin by determining the central conditions in the donor's progenitor at the time the CE phase begins.  The \citet{nele01a} data gives the progenitor's age and He core mass ($\Mti$) at this time.  These two quantities constrain the progenitor's initial mass. We use the EZ stellar evolution code \citep[derived from Peter Eggleton's stellar evolution code; see][and references therein]{paxton04} to calculate $\rho_c$, $T_c$ in a star of this initial mass and age.  The central conditions in the donor do not change during the rapid CE phase and we use the $\rho_c$, $T_c$ just found to determine the initial conditions for the WD cooling phase.  To calculate the WD cooling, we use the pure-He WD cooling models of \citet{alth97} (hereafter AB97). For a given WD mass, these models do not necessarily have a ($\rho_c$, $T_c$) pair matching that just found for the core (due to differences in the interior profile between the two models).  We map from the core model to the WD cooling model by fixing $\rho_c$ and $\Mti$, which then determines $T_c$ for the cooling model. While this is an ad hoc methodology, without being able to calculate the thermal evolution of the donor relaxing from its post-CE configuration to the WD configuration it is difficult to argue for a more accurate prescription.  In any case, the change in central specific entropy involved in this mapping is usually small compared to the entropy lost during the cooling.  We then use $\tcont$ and the AB97 cooling sequences to determine the donor's central conditions at contact. Finally, to determine the starting AM CVn evolution model, we find the DB03 model with $M_2 = \Mti$ and a specific entropy that equals the central specific entropy of the AB97 model at contact.  Generally speaking, most donors make contact as mildly to extremely degenerate objects, but there is a fraction of non-degenerate donors in the population.  Once the initial model is determined, we calculate $a$ from the requirement that $\RL=R_2$ at contact and integrate the system's evolution with $\Mdot$ calculated from equation (\ref{eq:mdot}).  We calculate $\Porb$ from equation (\ref{eq:pdrel}) and $n$ from our models assuming adiabatic evolution{.}

While providing a reasonable estimate of the central conditions in the donor's core, this prescription does not allow us to determine the evolution of the donor's outer regions prior to contact.  This will impact the evolution of these AM CVn systems at short $\Porb$.  For example, $R_2$ (at fixed $T_c$ and $M_2$) is smaller in our isentropic models than in the thermal equilibrium AB97 models, as is expected.  If we were to take the AB97 model's $R_2$ at contact instead of the DB03 $R_2$, the initial $\Porb$ would increase by up to a factor of 2-3, depending on the donor's degeneracy.  Additionally, using isentropic donors post-contact produces a different evolution than obtained by self-consistently evolving the structure of the donor at contact (something we are not able to do with our current models).  The discrepancy between the two evolution tracks should grow smaller as the donor loses mass and only the central regions of the star matter. 

Other factors also create uncertainty in the evolution during the early AM CVn phase. The cooling rate of a He WD depends on the amount of H left in its envelope after the CE event and whether H/He diffusive mixing is considered or not \citep{driebe99,alth01,alth97,benv98,sarna00}. The amount of H that remains after the CE is uncertain \citep[see, e.g.,][]{iben93}. In general, more H leads to slower WD cooling due to its greater opacity and the possible residual H burning in the lower atmosphere.  Further uncertainty is added by the questions of whether strong shell flashes can occur \citep{driebe99,sarna00,alth01}, altering the subsequent cooling rates by rapidly consuming large amounts of H.  All of this leads to significant uncertainty in the AM CVn population distribution at short $\Porb$.  However, pure He WDs cool the fastest and the isentropic donors have the most compact configuration for a given central specific entropy. Therefore our results provide a lower limit on the impact of finite $T_c$ on the WD-channel AM CVn population.

The uncertainties associated with the CE phase and the subsequent He WD cooling phase lead to uncertainties in $R_2$ near contact.  While this clearly impacts the short period AM CVn distribution, it may also be significant to the overall population. Donors with larger $R_2$ at contact---as expected for hotter, isothermal donors---are not as likely to experience a mass transfer instability due to advection of angular momentum onto the accretor \citep[either during a direct-impact accretion phase or because the accretion disk at short periods has a small radial dynamic range;][]{nele01a,marsh04}. Taking account of this effect could, therefore, increase the number of WD-channel systems that survive to become AM CVn binaries (as well as the upper limit on $M_2$ in those that do).

\subsection{The Resulting Population \label{sec:popdist}}  
We now compare the WD-channel population calculated using the DB03 $T=0$ \MR relations with the one calculated taking into account the pre-contact donor cooling.  We will refer to these two populations as the $T=0$ and RWDC (realistic WD cooling) populations, respectively. This $T=0$ population differs slightly from the analogous results of \citet{nele01a}, as their fully degenerate \MR relation differs from DB03. In the $M_2$ range of interest, the \citet{nele01a} donors are more compact than DB03's so that the resulting $\Mdot(\Porb)$ relation from the DB03 donors is shifted upward by $\approx 10-20\%$ from that for the \citet{nele01a} donors.  In the RWDC population, the track each system follows is determined by the donor's specific entropy at contact, with the relative number of systems evolving along each adiabat determined by the distribution in $\tcont$.  In the \citet{nele01a} data $\approx$ 10\%, 30\%, and 70\%  of the systems have $\tcont \leq$ 100 Myr, 500 Myr, and 1 Gyr, respectively, producing a significant fraction of systems not evolving along the $T=0$ track.  Since these numbers are based on using $T=0$ radii for the pre-contact donor, they are a lower limits to a self-consistently determined distribution of $\tcont$, where more systems would have hotter donors at contact.  The distribution of $\tcont$ also depends on the assumptions made in calculating the CE phase. A larger CE efficiency parameter (the ratio of the change in envelope binding energy to changes in orbital energy) than used by \citet{nele01a}, for example, would produce systems with wider separations at the end of the CE phase and reduce the number of systems with hot donors. 

\begin{figure}
\plotone{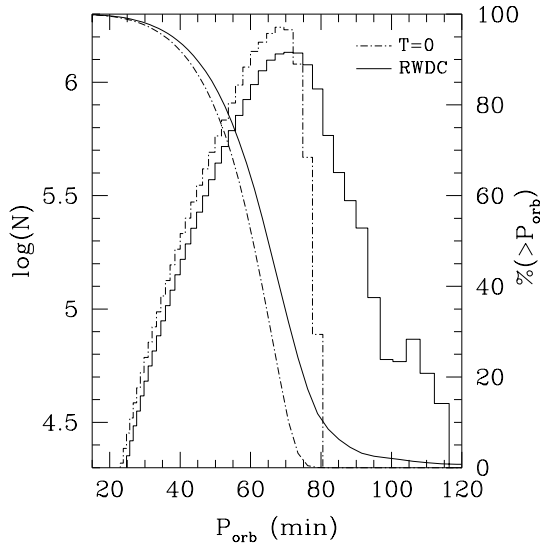}
\caption{A histogram of the number of systems as a function of $\Porb$ between the $T=0$ (dot-dashed line) and RWDC (solid lines) populations.  The higher specific entropy of some donors in the RWDC population leaks systems towards longer orbital periods as compared to the $T=0$ population.  The smooth curves show the percentage of each population lying above a given $\Porb$.  In the RWDC population, 65\% of systems have $\Porb>60$ min (50\% of the $T=0$ population lies above this period); 10\% of the RWDC population lies above the $T=0$ population's $\Porb\approx 80$ min cutoff.%
\label{fig:phist}}
\end{figure}

Figure \ref{fig:phist} shows the distribution of AM CVn binaries versus $\Porb$ in the two populations.  Both exhibit, at first, the rapid increase in the number of systems with increasing $\Porb$ due to the increase in binary evolution time with $\Porb$.  The $T=0$ population (dot-dashed lines in Figure \ref{fig:phist}) has a sharp cut-off at $\Porb \approx 80$ min set by how far a $T=0$ donor can evolve in 10 Gyr. Hot donors make contact at, and evolve out to, longer $\Porb$ than fully degenerate ones.  The range in the donor's specific entropy in the RWDC population therefore prevents a sharp cut-off in systems above $\Porb \approx 80$ min.  Instead there is a factor of $\approx 20$ decrease between the number of systems at $\Porb = 80$ and the number at $\Porb = 100$ min in the RWDC population.  The overall result is a shifting of systems from shorter to longer $\Porb$ in the RWDC population as compared to the $T=0$ population, resulting in a reduction in the number of systems we expect to see in the $\Porb$ range of the known AM CVn population (i.e. $\Porb < 65$ min).   In the RWDC population, 65 \% of systems have $\Porb>60$ min, while 50\% of the $T=0$ population lies above this period; 10\% of the RWDC population lies above the $T=0$ population's $\Porb = 80$ min cut-off.

\begin{figure}
\plotone{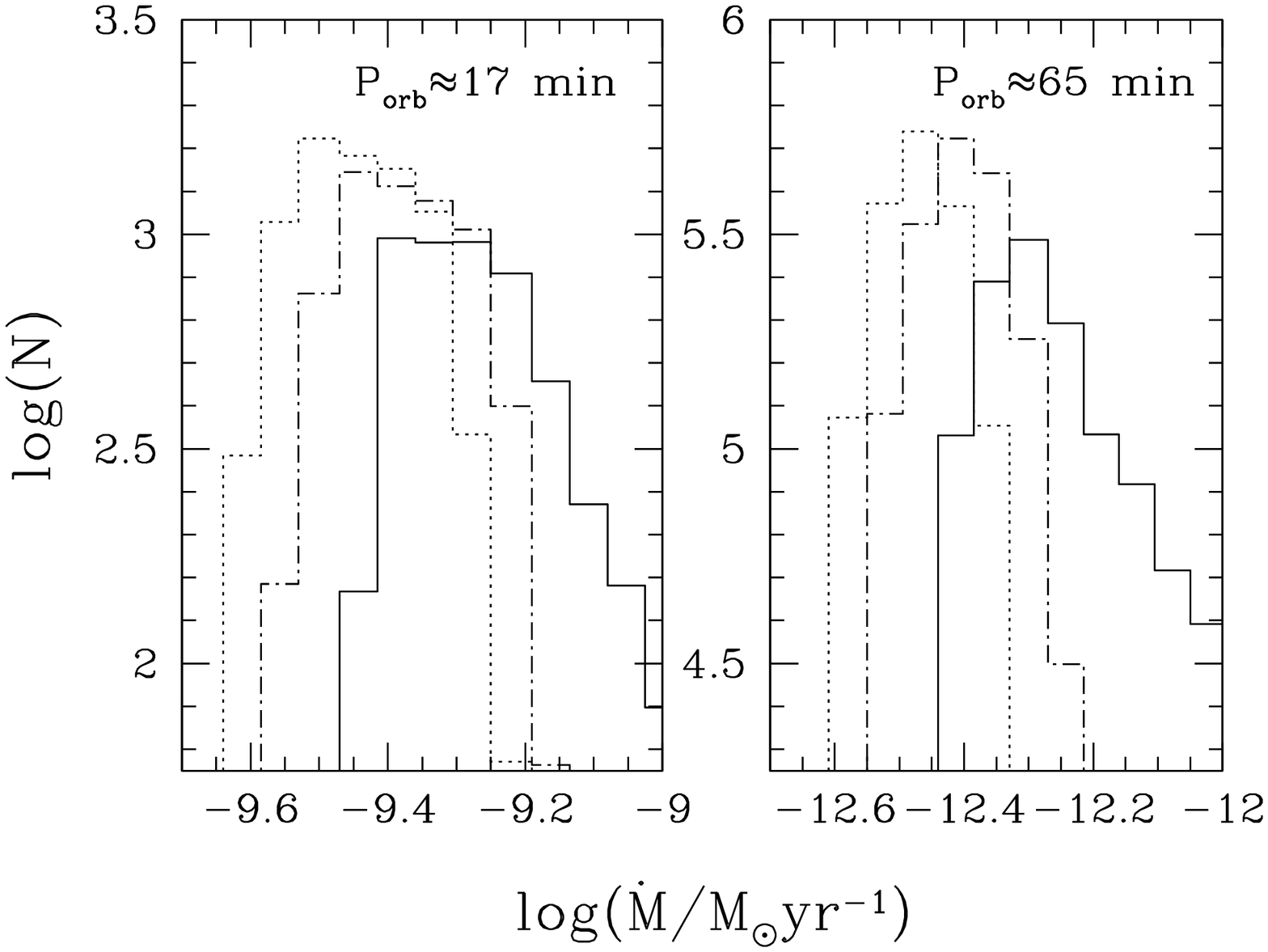}
\caption{{The distribution of systems in $\Mdot$ at the approximate $\Porb$ of AM CVn and CE-315.  Our RWDC population is shown by the solid line, the $T=0$ population is shown by the dot-dashed line, and the \citet{nele01a} $T=0$ results by the dotted lines.  A comparison between the \citet{nele01a} population and our $T=0$ population shows the magnitude of the shift in the $\Mdot$ distribution produced by the difference between the \citet{zapo69} \MR relation and our, more accurate, $T=0$ relations. The existence of hotter donors in the RWDC produces the tail seen in that population's $\Mdot$ distribution. At the longer $\Porb$ of CE-315, the spread in $\Mdot$ increases as specific entropy differences have more a significant impact on lower-mass donor's structure.}}
\label{fig:mdhist1}
\end{figure}

The range in $\Mdot$ that AM CVn binaries can attain at fixed $\Porb$ is increased in the RWDC population.  Figure \ref{fig:hemr} shows that at fixed $\Porb$, hotter donors must be more massive.  As $\Mdot \propto M_2^{2/3}\,M_2^2$ (at fixed $\Porb$ and when $M_2 \ll M_1$), hotter donors lead to a larger $\Mdot$ {(see also Figure \ref{fig:mdotp_isoT})}.  We show in Figure \ref{fig:mdhist1}  the distribution of the AM CVn WD-channel population $\Mdot$'s at the orbital periods of AM CVn, and CE-315 in the $T=0$ and RWDC populations.  In addition, we also show the results obtained with the \citet{nele01a} $T=0$ \MR relation to indicate the magnitude of the shift in $\Mdot$ produced when using our more accurate $T=0$ relation. The $\Mdot$ range in the $T=0$ populations is due to the range in $M_1$ at each $\Porb$. In the RWDC population, the existence of higher entropy donors leads to systems with larger $M_2$ at a given $\Porb$ and the tail towards higher $\Mdot$ seen in this distribution.  The offset in the minimum $\Mdot$ seen between the $T=0$ and the RWDC population  {results from the fact that for the lowest accretor masses in the RWDC population, none have donor's that are close to fully degenerate}.  It is unclear if this is a robust result or simply a consequence of the particular population synthesis calculation used here to provide the initial conditions.  The number of systems in the tail of the distribution increases with $\Porb$, as a higher specific entropy has a more significant impact on the donor's structure at lower $M_2$.  The majority of systems in the RWDC population lie near the $T=0$ tracks as is expected, but $\sim 10\%$ of this population have donors that produce mass transfer rates greater than the spread attributable to $M_1$ variations alone (at least at longer $\Porb$; {see the right hand panel in} Figure \ref{fig:mdhist1} in particular).

\subsection{Implications for the AM CVn Gravity Wave Signal \label{sec:gwsig}}
The galactic population of AM CVn binaries contains objects that  {can be detected with} the planned {\it Laser Interferometer Space Antenna} (LISA) gravity wave detector mission (see http://lisa.nasa.gov/ and http://sci.esa.int/home/lisa/ for mission details), and the gravity wave signal of the AM CVn binary population has been considered in several recent studies \citep{nele01c,farm03,nelemans04}.  Here we discuss the differences between the $T=0$ and RWDC populations' gravity wave signals.

\begin{figure}
\plotone{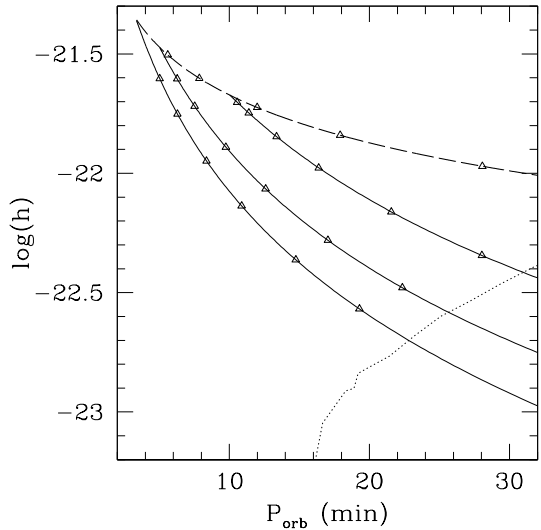}
\caption{An AM CVn binary's (with $\Moi = 0.6 \msun$, $\Mti = 0.2 \msun$ at 1 kpc) $h$ versus $\Porb$, showing the impact of a hot donor on the system's post-contact GW flux.  The dashed line shows the system's pre-contact inward evolution, the symbols showing when its time to contact (for a $T=0$ donor) is $\log(t/\mathrm{yr}) = 5.0,\,5.5,\,6.0,\,6.5,$ and $7.0$, left to right.  The solid lines show the post-contact evolution for a $T=0$ donor and for donors hot enough to make contact at $\Porb =5\,(10)$ min.  The symbols here show time since contact: $\log(t/\mathrm{yr}) = 5.0,\,5.5,\,6.0,\,6.5,\,7.0$ and $7.5$, left to right.  At fixed $\Porb$, hotter donors are more massive and louder GW sources.  The dotted line shows the detached WD-WD binary confusion limit for LISA \citep{nele01c}.
\label{fig:hfdiff}}
\end{figure}

A binary system in a circular orbit emits gravity waves at a frequency $f=2/\Porb$ (i.e. twice the orbital frequency) and luminosity, $L_{\mathrm{GW}}$, \citep{press72}
\begin{equation}
L_{\mathrm{GW}} = \frac{32}{5} \frac{G^4}{c^5} \frac{M_1^2 M_2^2 (M_1+M_2)}{a^5}\,,
\label{eq:lgw}
\end{equation}
where $G$ is the gravitational constant and $c$ the speed of light.  The flux, $F=L_{\mathrm{GW}}/4 \pi d^2$, received by a detector a distance, $d$, from the source is often written in terms of the so-called dimensionless strain amplitude, $h$, given by \citep{press72,nele01c}
\begin{equation}
\begin{split}
h &= \left[\frac{4 G}{c^3 \pi f^2} F\right]^{1/2}\\
  &=\ee{5.0}{-22}\left(\frac{\mathcal{M}}{\msun}\right)^{5/3} \left(\frac{\Porb}{\mathrm{hr}}\right)^{-2/3} \left(\frac{d}{\mathrm{kpc}}\right)^{-1}\,,
\end{split}
\label{eq:gwh}
\end{equation}  
where $\mathcal{M}=(M_1 M_2)^{3/5}/(M_1+M_2)^{1/5}$ is known as the chirp mass.  In an individual AM CVn binary, changing the entropy of the donor changes the relation between $\mathcal{M}$ and $\Porb$ and alters the binary's GW luminosity.  We show a summary of how hot donors change the expected GW signal in Figure \ref{fig:hfdiff},  {where we} consider the evolution of an AM CVn binary with $\Moi=0.6 \msun$ and $\Mti = 0.2 \msun$ located 1 kpc away. The evolution of this binary pre-contact is independent of the donor's state{,} and the system's $h$ as a function of $\Porb$  {evolves inward along} the dashed line in Figure \ref{fig:hfdiff}  to{ward} shorter periods.  If the donor is fully degenerate, the binary makes contact at $\Porb \approx 3.5$ min and then evolves outward in $\Porb$ along the left-most solid line.  The orbital period at contact depends on the donor's entropy.  The other two solid lines show the evolution for systems making contact at $\Porb = 5$ min (requiring a partially degenerate donor with $T_c = \ee{5.6}{7}$ K at contact) and $10$ min (requiring a non-degenerate donor with $T_c \approx \ee{6.1}{7}$ K at contact), respectively.  Donors with these entropies occur in the RWDC population {(}which contains systems that make contact at orbital periods as large as $\approx 25$ min{)}.  The qualitative impact of a hot donor is the increase in $h$ as a function of $\Porb$ as hotter donors  at fixed $\Porb$ are more massive, increasing $\mathcal{M}$. 

{The integrated GW flux from a collection of identical AM CVn binaries depends on the donor's specific entropy in two ways.  First, the flux from each system at a fixed $f$ increases with the donor's specific entropy. Second, each system's $\dot{f} = df/dt$ depends on $n$, which varies with both the donor's mass and specific entropy. For constant $n$ in a steady-state population, one can derive a simple scaling for the GW energy density per logarithmic frequency interval, $\Edgw \propto f N(f) \Lgw(f)$ $\propto f\,\Lgw(f)/\dot{f}$ $=f \,d \Egw/df$. This can be expressed in terms of the dimensionless characteristic amplitude in a logarithmic interval, $h_c$, as 
\begin{equation}
\Edgw = \frac{\pi f^2 c^2}{4 G} h_c^2(f)\,. 
\end{equation}
Taking $R_2 \propto M_2^n$, and assuming $M_2 \ll M_1$ along with stable mass transfer, $h_c^2 \propto f^\alpha$ where
\begin{equation}
\alpha = \frac{2+12n}{3(1-3n)}\,,
\label{eq:hcan}
\end{equation}  
\citep[see][for a derivation of this under the assumption $n=-1/3$]{phinney01,farm03}.  Returning to Figure \ref{fig:hfdiff}, at contact $n=-0.350$ for the fully degenerate donor, while for the donor making contact at 5 min (10 min) , $n=-0.342$ (-0.333).  At $\Porb = 20$ min ($f = \ee{1.67}{-3}$ Hz), the three donors have $n=-0.233,\,-0.276,$ and $-0.308$ respectively.  As $\alpha$ is an increasing function of $n$ and is negative for $n<-0.167$,  $h_c^2$ falls off less rapidly with $f$ for cold donors (as they evolve less rapidly in $f$ than donors with higher specific entropy). While hotter donors produce AM CVn binaries that are individually more GW luminous at fixed $f$, their increased rate of $f$ evolution somewhat mitigates this in the population's overall GW energy density at lower $f$.}

\begin{figure}
\plotone{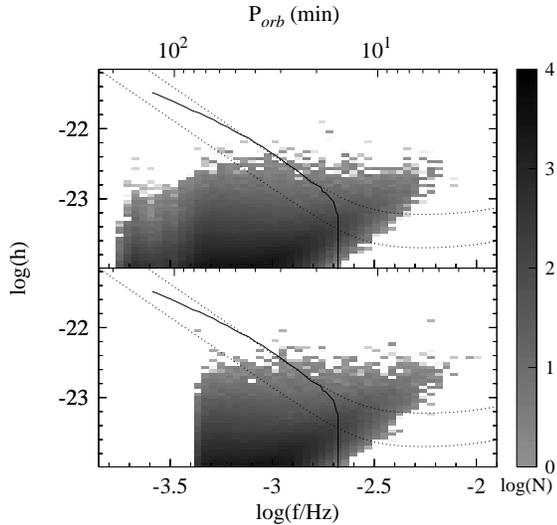}
\caption{ The distribution in the galactic population of WD-channel AM CVn binaries in $h$ and $f$ {for the RWDC (upper panel) and $T=0$ (lower panel) populations}. The grey-scale indicates the logarithm of total number of sources in each $(f,\,h)$ bin.  The solid line shows the WD-WD confusion limit calculated by \citet{nele01c}, while the dotted line{s} shows LISA's predicted sensitivity for a S/N ratio of 1 {(lower line) and 3 (upper line)} \citep{lars00,lars02}. 
\label{fig:fhhist}}
\end{figure}

We now move from these simple analytics to a direct numeric calculation of the GW signal from both the $T=0$ and RWDC populations.  We begin, following \citet{nele01c}, by randomly distributing the systems in each population in galactic coordinates with uniform probability in galactic azimuth and with a probability, $P(R_G,\,z)$, in galactic radius, $R_G$, and height above the mid-plane, $z$ of 
\begin{equation}
P(R_G,\,z) = \frac{1}{2 H^2 \beta} \exp\left[-\frac{R_G}{H}\right] \mathrm{sech} \left(\frac{z}{\beta}\right)^2 \,,
\end{equation}
where $\beta=200$ pc and $H=2.5$ kpc.  To calculate $d$ for each system we took the position of the sun to be $R_{\mathrm{sun}}=8.5$ kpc and $z_\mathrm{sun}=-30$ pc, again following \citet{nele01c}. We then calculated $h$ for each system from its $d$ and current orbital parameters using equation (\ref{eq:gwh}).  Figure \ref{fig:fhhist} presents the resulting distribution of AM CVn binaries in $f$ and $h$ for each of these populations. As expected, the qualitative difference between the $T=0$ (lower panel) and RWDC (upper panel) distributions is a shift in individual systems to lower $f$ in the RWDC population with an extended tail in the RWDC distribution to frequencies lower than $f \approx \ee{4.0}{-4}$ Hz.   

The fraction of this AM CVn population that LISA will detect is set by LISA's sensitivity and the galactic population of detached WD-WD binaries. For a signal to noise (S/N) ratios of 1 and 3, LISA's predicted sensitivity \citep[as calculated with Shane Larson's online generation script, http://www.srl.caltech.edu/~shane/sensitivity/, based on the results of][]{lars00,lars02} is shown by the dotted lines in Figure \ref{fig:fhhist}. Population synthesis studies indicate that detached WD-WD binaries will dominate the GW signal at frequencies $f \lesssim 10^{-3}$ Hz \citep{evans87,lipunov87,hils90,nele01c,farm03}. These systems are expected to produce a so-called ``confusion limit'', a noise background below $\approx 2$ mHz produced by the vast number of systems that will not be individually resolvable.  The solid lines in Figure \ref{fig:fhhist} shows the confusion limited noise set by this background (essentially the averaged GW signal from detached WD-WD binaries) as calculated by \citet{nele01c}.  The WD-channel AM CVns in the $T=0$ population that will be detectable above both the LISA sensitivity curve and the confusion limited noise have an average $d$ of 8.6 kpc.  The corresponding RWDC population contains  $\approx 13\%$ fewer systems at an average $d$ of 8.7 kpc. The decrease in the RWDC numbers is due mainly to the shifting of systems towards shorter $f$.  Even though individual RWDC systems are on average louder than the $T=0$ systems, the volume over which LISA will detect these systems is not significantly altered. This is due to the fact most of the AM CVn systems above the confusion limit in either population are detectable to the edge of the galaxy (the galactic geometry sets the lower right cutoff of both distributions in Figure \ref{fig:fhhist}). 

\begin{figure}
\plotone{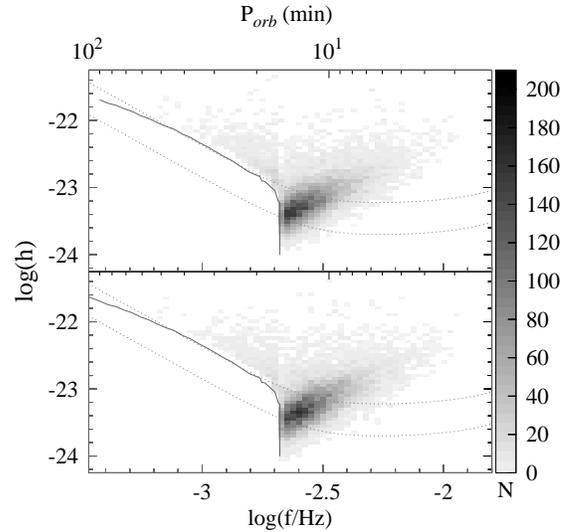}
\caption{The distribution of LISA \emph{resolved} systems in the RWDC (upper panel) and $T=0$ (lower panel) populations.  The grey-scale indicates the total number of sources in each $(f,\,h)$ bin.  The solid gray line shows the confusion limit calculated by \citet{nele01c}, while the dotted line shows LISA's predicted sensitivity for a S/N ratio of 1 {(lower line) and 3 (upper line)} \citep{lars00,lars02}.    %
\label{fig:fhhist_res}}
\end{figure}

\begin{figure}
\plotone{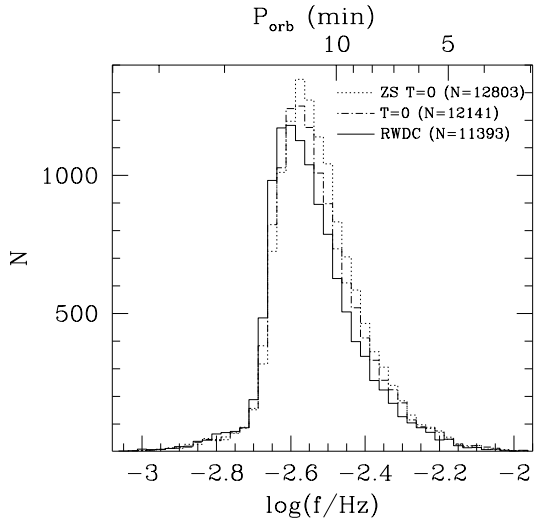}
\caption{A histogram of LISA resolved  sources (assuming a S/N of 1) versus $f$ for the $T=0$ population (dot-dashed line) and the RWDC population (solid line). For comparison, the distribution obtained for the $T=0$ population when using the \citet{nele01a} degenerate \MR relation is shown by the dotted line. All distributions cut-off at the 2 mHz confusion noise cutoff.  The RWDC distribution's mean frequency is shifted downward $\approx 10^{-4}$ Hz from that of the DB03 $T=0$ distribution. The resolved RWDC population has $\approx 6\%$ fewer systems due to the increased number of systems that evolve below 2 mHz.  The RWDC population has 11\% fewer resolved systems than the \citet{nele01a} $T=0$ population.
\label{fig:ghhist}}
\end{figure}

For an estimated mission length of 1 yr, LISA will be able to resolve frequencies separated by $\Delta f \approx 1/1\,\mathrm{yr} = \ee{3}{-8}$ Hz.  AM CVn binaries with an $h$ above the confusion limited noise and that are the only system in a given resolved frequency bin will be individually resolved by LISA (this is a bit of an oversimplification since other populations, such as detached WD-WD binaries, not included in our calculation will also contribute to the LISA signal; as we are interested here in how a range in donor specific entropy impacts the WD-channel population, we ignore this added complication).  Between the $T=0$ and RWDC populations, the distribution of LISA \emph{resolvable} sources differs slightly, as we show in Figure \ref{fig:fhhist_res}. The main difference between the two distributions is the slight shift in the maximum of the RWDC population relative to that of the $T=0$ and a decrease in the total number of resolved systems.  This is seen more clearly in Figure \ref{fig:ghhist} were we show the histogram of resolved systems in each population versus $f$.  Overall, the RWDC population contains $\approx 6 \%$ fewer resolved systems than the $T=0$ and the RWDC distribution's mean frequency is $\approx 10^{-4}$ Hz less than the $T=0$'s. As compared to the results for the fully degenerate distribution calculated using the \cite{nele01a} degenerate \MR relations, the RWDC contains $\approx 11\%$ fewer systems. 

\subsection{Future Calculation Refinements}
These population calculations are only a first step towards a more realistic theoretical treatment of the AM CVn population.  Here we speculate on the possible results of a more refined calculation. In particular, how would improving on the approximations necessitated by our current donor models alter our result?  The choice of adiabatic evolution assumes that the donor cools on a time scale long compared to the mass loss time scale ($\sim M_2/\Mdot$).  While this is almost certainly the case for systems near contact, it may not be the case for systems at long $\Porb$.  If and when adiabaticity is violated depends on how the donor's luminosity scales with $T_c$ and $M_2$.  If adiabatic evolution doesn't hold, then some fraction of the long $\Porb$ systems in the RWDC population could drop out of the AM CVn population as the donor cools and contracts within its Roche Lobe, reducing the number of expected long period AM CVn systems. Whether GW losses can subsequently bring such systems back into contact will depend on how much the donor contracted to its fully degenerate configuration and on the binary's parameters.  The expectation would be that the orbit at this point would evolve on time scales of several Gyr or longer, so it would be doubtful that systems falling out of contact would reestablish contact later.  The details of such scenarios remains to be worked out with a more sophisticated calculation.

\section{Summary and Discussion \label{sec:conclusions}}
We have applied the DB03 low-mass, isentropic He WD models, originally calculated for the donors in ultracompact low-mass X-ray binaries to the AM CVn population.  This model set provides a continuous relationship between a model's $M_2$, $R_2$ and its specific entropy,  allowing us to consider the impact of donor specific entropy on the evolution of interacting binaries.  Higher entropy donors have larger $R_2$ and, for a given $\Porb$, are more massive than $T=0$ objects, producing a higher $\Mdot$.  Across a population of AM CVn systems, variations in the donor's specific entropy produces a range of \MR relations, leading to a range in possible $M_2$ and $\Mdot$ at fixed $\Porb$.  For a reasonable range in $T_c$, this range in $\Mdot$ can be up to an order of magnitude or so (Figure \ref{fig:mdotp_adiab}).

This also means that there is a range in parameters allowed {for} the known AM CVn systems.  Most systems have a measured (or inferred) $q$, which, with the measured $\Porb$, we used to determine a unique $M_2$-$\Mdot$ relation for each system.  Structural arguments (the donor cannot be more compact than a $T=0$ object nor the accretor more massive than 1.4 $\msun$) and the photometric behavior of each system allow us to restrict the range in $M_2$.  The limits we derived allow most systems a factor $\sim$ 3 range in $\Mdot$. One possible exception is V803 Cen, which, with its extremely low, but uncertain $q$ \citep{kato04}, is indicated to have a donor of $M_2 \approx 0.02 \msun$ and an accretor with a minimum mass of 1.3 $\msun$; if this is truly the case, its $\Mdot \approx 10^{-10} \msun$ yr$^{-1}$.  Our donor models also allow us to constrain each donor's $T_c$ range and correlate $T_c$ with $M_2$, \emph{providing the first means of inferring internal donor properties in the AM CVn systems}.  A summary of the these relations between $M_2$, $\Mdot$, and $T_c$ are shown in Figure \ref{fig:mdot_m2} and listed in Tables \ref{tab:systems} and \ref{tab:systems_disk}.

Across the galactic AM CVn population, variations in individual system's evolution will leave an imprint on the overall population distribution.  Our model set is ideally suited to explore how variations in the time between leaving the CE phase and initiating mass transfer in the AM CVn phase impacts the $\Mdot$-$\Porb$ distribution of the WD-channel AM CVn systems.  Starting with the data for the WD-channel from the \citet{nele01a} population synthesis model, we determined each donor's central specific entropy using the EZ stellar evolution code \citep{paxton04} to determine the donor's initial conditions at the end of the CE phase. We then used each donor's $\tcont$ and the \citet{alth97} He WD models to determine how much each donor cooled by the time the AM CVn phase started. We evolved each system adiabatically using the DB03 models to determine the present day orbital configuration for each system in the population. The results of this calculation are approximate for several reasons. Calculating the CE phase and He WD cooling involves several uncertainties \citep{driebe99,sarna00,alth97,benv98,alth01}.  The use of pure He models, which cool faster than He WD models with H present in their envelopes, means our results give a lower limit on the initial donor entropy in the WD-channel systems.

The DB03 models have an isentropic interior profile.  At contact, most donors should have  close to an isothermal core and an entropy profile that increases outward.  At short $\Porb$, the \MR relations of the DB03 models are therefore likely not a good approximation for actual donors;  the use of isentropic models becomes increasingly better as the donor's mass is reduced and the entropy difference between its core and outer boundary is reduced.  Therefore, the $\Porb$ distribution $\lesssim 20$ min calculated here is subject to uncertainty.  At periods greater than this, $M_2 \lesssim 0.1 \Mti$ and isentropic models provides a good approximation to the actual donors that becomes increasingly better at lower $M_2$ and longer $\Porb$.  

 We compare our conservative calculation that includes WD cooling (the RWDC population) to the same population evolved along the $T=0$ track.  The RWDC $\Porb$ distribution shows slightly fewer systems at short $\Porb$ compared to the $T=0$ and has an extended tail at $\Porb>80$ min, the period at which the $T=0$ population cuts off due to the finite age of the galaxy; 65\% of the RWDC population has $\Porb > 60$ min, compared to 50\% of the $T=0$ and 10\% of the RWDC population has $\Porb > 80$ min.  At fixed $\Porb$, the $\Mdot$ distribution of the RWDC population has a tail towards higher $\Mdot$ compared to the $T=0$ population.  The fraction of systems with $\Mdot$ higher than the $T=0$ population increases with $\Porb$ since a higher specific entropy impacts the structure of low-mass donors more significantly.  In addition to the differences between these two populations,  there is a difference between the \citet{nele01a} $T=0$ population and the $T=0$ calculated with the DB03 models.  In the $M_2$ range of interest, the radii of the \citet{nele01a} donors  {are} less than the DB03, introducing a shift in the distribution of $\Mdot$ at fixed $\Porb$ between the two fully degenerate populations;  the DB03 $T=0$ $\Mdot$ distributions are shifted a factor of $\approx 1.15-1.5$ above the \citet{nele01a} $T=0$ distributions. 
We also calculated the gravity wave (GW) signal the RWDC and our $T=0$ populations would produce, assuming the systems were distributed throughout the galaxy according to the distribution used by \cite{nele01c}. The RWDC population contains 13 \% fewer systems that LISA will be sensitive enough to detect (assuming a S/N of 1) with a large enough $h$ to be seen above the unresolved background of detached WD-WD binaries.  This is due to the increased number of systems in the RWDC population evolving out to longer $\Porb$ and being lost under the confusion noise.  In terms of systems that LISA will individually resolve, the RWDC population has 6\% fewer systems than the DB03 $T=0$ population and 11\% fewer than the \citet{nele01a} population.     

One question posed to the theory community by the AM CVn population is why there are so few systems known.  Currently, the most optimistic estimate of the local density of these systems is a factor of 5 below the most pessimistic estimates from theoretical population synthesis models \citep{nele01a,groot03}.  The question is whether this is due to overall normalization problems in the population models, incompleteness of the known sample, a results of uncertain physics involved in binary evolutionary (such as the CE phase),  or if there is unexplored physics involving the donor or the accretor that remove systems from the AM CVn population or caused them to appear differently from what we expect. Further theoretical investigations of this latter possibility, along both the lines discussed above and others, will help answer these questions and be useful in constraining uncertainties in the physics governing the evolution of this, and other, binary stellar populations.

\acknowledgments
We would like to thank Joe Patterson, Boris G\"{a}nsicke, Danny Steeghs, and Tom Prince for their technical input and encouragement over the course of preparing this manuscript, as well as Jan-Erik Solheim and  Brian Warner for several discussions.  {We also thank the anonymous referee for their thorough review and constructive suggestions that improved our discussion}.  This work was supported by the National Science Foundation under grants PHY 99-07949, AST02-05956, and by NASA through grant AR-09517.01-A from STScI, which is operated by AURA, Inc, under NASA contract NAS5-26555.

\end{document}